\newcommand{\CLASSINPUTtoptextmargin}{2cm}
\newcommand{\CLASSINPUTbottomtextmargin}{2.5cm}
\documentclass[conference]{IEEEtran}
\IEEEoverridecommandlockouts

\usepackage[dvipdfmx]{graphicx}
 \usepackage{lipsum,graphicx,subcaption}
\usepackage{float}
\usepackage[font=small]{caption}
\usepackage[labelformat=simple]{subcaption}

\usepackage{cite}
\usepackage{amsmath,amssymb,amsfonts,amsbsy}
\usepackage{algorithmic}
\usepackage{textcomp}
\usepackage{algorithmic,algorithm}
\usepackage{url}
\usepackage{color}
\usepackage[11pt]{moresize}
 
\def\BibTeX{{\rm B\kern-.05em{\sc i\kern-.025em b}\kern-.08em
    T\kern-.1667em\lower.7ex\hbox{E}\kern-.125emX}}

\newcommand{\rmnum}[1]{\romannumeral #1}
\usepackage{mathtools}
\usepackage{amsthm}
\theoremstyle{definition}

\usepackage{mathabx}
\usepackage{optidef}
\usepackage{enumitem}

\usepackage{tabulary}
\usepackage{hhline}

\usepackage{xcolor}
\usepackage{listings}
\usepackage{hyperref}

\definecolor{mGreen}{rgb}{0,0.6,0}
\definecolor{mGray}{rgb}{0.5,0.5,0.5}
\definecolor{mPurple}{rgb}{0.58,0,0.82}
\definecolor{backgroundColour}{rgb}{0.95,0.95,0.92}

\lstdefinestyle{CStyle}{
    backgroundcolor=\color{backgroundColour},   
    commentstyle=\color{mGreen},
    keywordstyle=\color{magenta},
    numberstyle=\color{mGray},
    stringstyle=\color{mPurple},
    basicstyle=\footnotesize,
    breakatwhitespace=false,         
    breaklines=true,                 
    captionpos=b,                    
    keepspaces=false,                 
    numbers=left,                    
    numbersep=5pt,                  
    showspaces=false,                
    showstringspaces=false,
    showtabs=false,                 
    tabsize=2,
    language=C++
}

\begin{document}
\title{\Huge Spatial Indexing for System-Level Evaluation of 5G Heterogeneous Cellular Networks}


\author{
    \IEEEauthorblockN{Roohollah~Amiri\IEEEauthorrefmark{1}, Eren~Balevi\IEEEauthorrefmark{2}, Jeffrey~G.~Andrews\IEEEauthorrefmark{2}, Hani~Mehrpouyan\IEEEauthorrefmark{1}}\\
    \IEEEauthorblockA{\IEEEauthorrefmark{1}{Department of Electrical and Computer Engineering, Boise State University, ID 83725 USA}}
    \IEEEauthorblockA{\IEEEauthorrefmark{2}{Department of Electrical and Computer Engineering, The University of Texas at Austin, TX 78701 USA}}
    \IEEEauthorblockA{Email:~{roohollahamiri@u.boisestate.edu, erenbalevi@utexas.edu, jandrews@ece.utexas.edu, hanimehrpouyan@boisestate.edu}}
}

\maketitle

\begin{abstract}
System level simulations of large 5G networks are essential to evaluate and design algorithms related to network issues such as scheduling, mobility management, interference management, and cell planning. In this paper, we look back to the idea of spatial indexing and its advantages, applications, and future potentials in accelerating large 5G network simulations. We introduce a multi-level inheritance based architecture which is used to index all elements of a heterogeneous network (HetNet) on a single geometry tree. Then, we define spatial queries to accelerate searches in distance, azimuth, and elevation. We demonstrate that spatial indexing can accelerate location-based searches by $3$ orders of magnitude. Further, the proposed design is implemented as an open source platform freely available to all.
\end{abstract}

\section{Introduction}

Supporting the expected cellular traffic growth is one of the main tasks for the next generation (a.k.a ``5G'') wireless cellular networks and densification is one of the main technologies to achieve such growth~\cite{art_5Gwillbe}. A key driver for densification in the next $5$-$10$ years will be small base stations (SBSs) operating at millimeter wave (mmWave) frequencies. These SBSs will also support conventional communication below 6~GHz (Sub6GHz) frequencies and possibly use mmWave for the backhauling as well as some user equipment (UE) connections. Furthermore, due to propagation features in mmWave bands, usage of highly directional antennas at the transceivers is a necessity~\cite{art_Jeff_0}. Hence, 5G will contain directional heterogeneous networks (HetNets) with large number of nodes working on different frequency bands.

In the development and standardization of 5G, simulations are necessary to implement and design new algorithms and protocols. Considering the above features of 5G, system-level simulations need platforms which deliver accurate results in short time in large HetNets. These simulations are needed to evaluate the performance of scheduling algorithms, mobility managements procedures, interference management methods, and cell planning algorithms~\cite{art_sys_level_LTE}.

In simulation of large networks, operations that require searches over various nodes of the network may be extremely time consuming, where spatial indexing has been one of the methods to address this issue~\cite{book_Preparata}. In fact, spatial indexing has been used instead of traditional array indexing in order to accelerate location-based searches in the simulation of large homogeneous networks such as wireless sensor networks (WSNs). Wireless sensors are indexed based on their location on a geometry tree to provide fast search queries. This method can not be trivially applied in HetNets since a single geometry tree cannot be used for spatial indexing of different nodes. In this paper, first, we propose a multi-level inheritance based structure to be able to store different nodes of a HetNet on a single geometry tree. The proposed structure is polymorphic in a sense that different levels of a node can be accessed via dynamic casting~\cite{book_Stroustrup}. Second, we focus on potentials of spatial indexing in accelerating the simulation of directional communications. We introduce different spatial queries and show that spatial indexing significantly accelerates simulation time in orders of magnitude when it comes to location-based searches over azimuth, and elevation as well as its traditional usage in searches over distance.



\subsection{Motivation}

Traditional wireless network simulators such as Network Simulator (NS-2, NS-3)~\cite{art_ns, art_ns3} do not take into consideration the relationship between a terminal and its location in the indexing procedure. In other words, nodes are indexed based on features such as identification numbers or the order in which they are added to the network. Nodes are simply stored in an array (array indexing) and there is no pre-processing (sorting or classification) based on the location of the nodes. Consequently, in a network with $n$ nodes, the search size of any algorithm related to the location of the nodes equals the total number of the nodes in the network, i.e., $\mathcal{O}(n)$. Hence, if all nodes run such an algorithm, the exhaustive search complexity would be $\mathcal{O}(n^2)$. This is important since in dynamic wireless networks,  location-dependent searches are called frequently in simulations. Examples of location-dependent searches in a simulation environment are: finding the nearest neighboring users for association or handover purposes, finding the k-nearest neighboring BSs for coordinated multipoint (CoMP) transmission, or finding the potential interferers to evaluate signal-to-interference-plus-noise-ratio (SINR) of a communication link. While the above searches are defined over distance, in mmWave applications the direction of communication is important as well. This means searching over distance, azimuth, and elevation at the same time which can increase the complexity of the overall algorithm significantly. In practice, decreasing the order of search complexity can potentially change the computation time from hours to seconds of computation in large networks. In order to achieve this goal, location-transparency can be changed into location-dependency in the indexing of nodes~\cite{art_dunham, art_query_survey, art_arch}. To this aim, spatial indexing has been used in homogeneous networks with the intent of accelerating distance queries. In this work, we take advantage of polymorphic programming to use spatial indexing in heterogeneous cellular networks and to provide fast spatial search queries in distance, azimuth, and elevation.

\subsection{Related Works}

There are several open-source simulators developed for different purposes in wireless networks. In this category, with the focus on open-source platforms, we have Network Simulators (NS-2, NS-3)~\cite{art_ns, art_ns3}, OMNET++~\cite{art_omnet}, J-Sim~\cite{art_jsim}, and SHOX~\cite{art_shox} platforms. These common simulators focus on preparing a platform for design and evaluation of communication protocols in the network layer or layers above it. 
The physical layer modules in the above platforms are not appropriate for mmWave or directional communications. The full-stack mmWave module proposed by~\cite{art_End2End} alleviates this shortcoming, by adding this module to the NS-3 platform for support of the mmWave communications. The physical-layer module presented by~\cite{art_End2End} is extensive. However, the added module is built on the core of the NS-3 and is not designed to calculate directional interference in networks with dynamic topology. Nevertheless, none of the above simulators takes advantage of spatial indexing. In fact, the nodes of the network are simply stored in an array.

Spatial indexing has been used in two major applications in wireless communications: $(\rmnum{1})$ location-aware data indexing in wireless broadcast systems and $(\rmnum{2})$ location-dependent indexing in simulation platforms. Location awareness is naturally the first step of context-awareness in broadcast systems and spatial indexing is used in wireless broadcast systems where efficient indexing of data segments is essential for mobile users to find their required query from the broadcast data~\cite{art_broadcast0, art_broadcast1, art_broadcast2}. 
Apart from the application of spatial indexing in broadcast systems, its advantage in simulation environments has been noticed in a few works~\cite{report_sandia, art_vanet}. Fast distance query of spatial indexing is used in high fidelity large-scale simulation of wireless communications~\cite{report_sandia}. Also,~\cite{art_vanet} changed the indexing procedure of NS-3 in the simulation of a large-scale (more than $1000$ nodes) vehicular ad-hoc network (VANET) to provide fast distance queries as well. However, wireless networks and more specifically cellular networks are heterogeneous. This means elements of the network can vary from macro base stations to small base stations and mobile users. Further, there are certain phenomena that need to be considered in system-level simulations such as blockages. Also, in 5G, features of millimeter-wave communications such as directionality changes the complexity of location-dependent searches. In fact, search queries are not just in distance but also in azimuth and elevation as well. Therefore, we need an architecture that uses spatial indexing in a HetNet and supports the above features. In the following, we introduce a generic architecture that uses multi-level inheritance and polymorphism to enable indexing a heterogeneous network on a single geometry tree. Then we evaluate the performance of the proposed architecture with respect to traditional array indexing.


It is worth mentioning that acceleration of high-fidelity wireless communication simulations
has been investigated through geography-based partitioning for parallel implementation on multiple processors as well~\cite{art_fidelity}. However, parallel computing is not the focus of this work.


\section{Spatial Indexing and Queries in HetNets}

In this section, we first introduce the architecture which enables us to use spatial indexing for a HetNet. Then, the indexing procedure and the defined spatial queries are explained.






\begin{figure}
    \centering
    \includegraphics[width=1.\columnwidth]{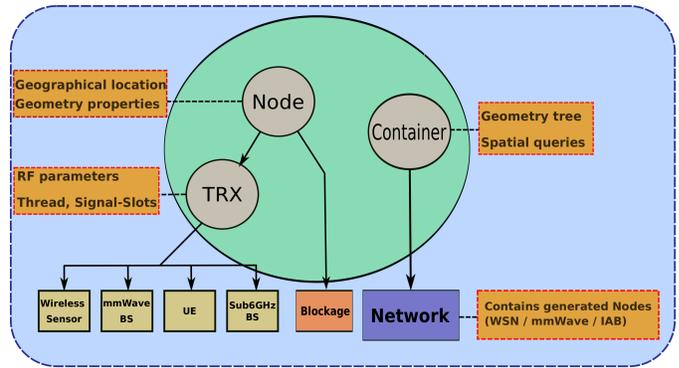}
    \caption{\small{The proposed multi-level architecture. Location and geometry properties of elements of the network are abstracted in the \textit{Node} object. The \textit{Container} stores the elements of the network on a geometry tree. From the view of the geometry tree, all elements are the same and represented as polygons. The Network basically represents any wireless network which can be a WSN, mmWave, or an integrated access and backhaul (IAB) network.}}
    \label{fig_hierarchy}
\end{figure}

\begin{figure*}
    \centering
    \begin{subfigure}[t]{0.7\columnwidth}
        \centering
        \includegraphics[width=1.0\columnwidth]{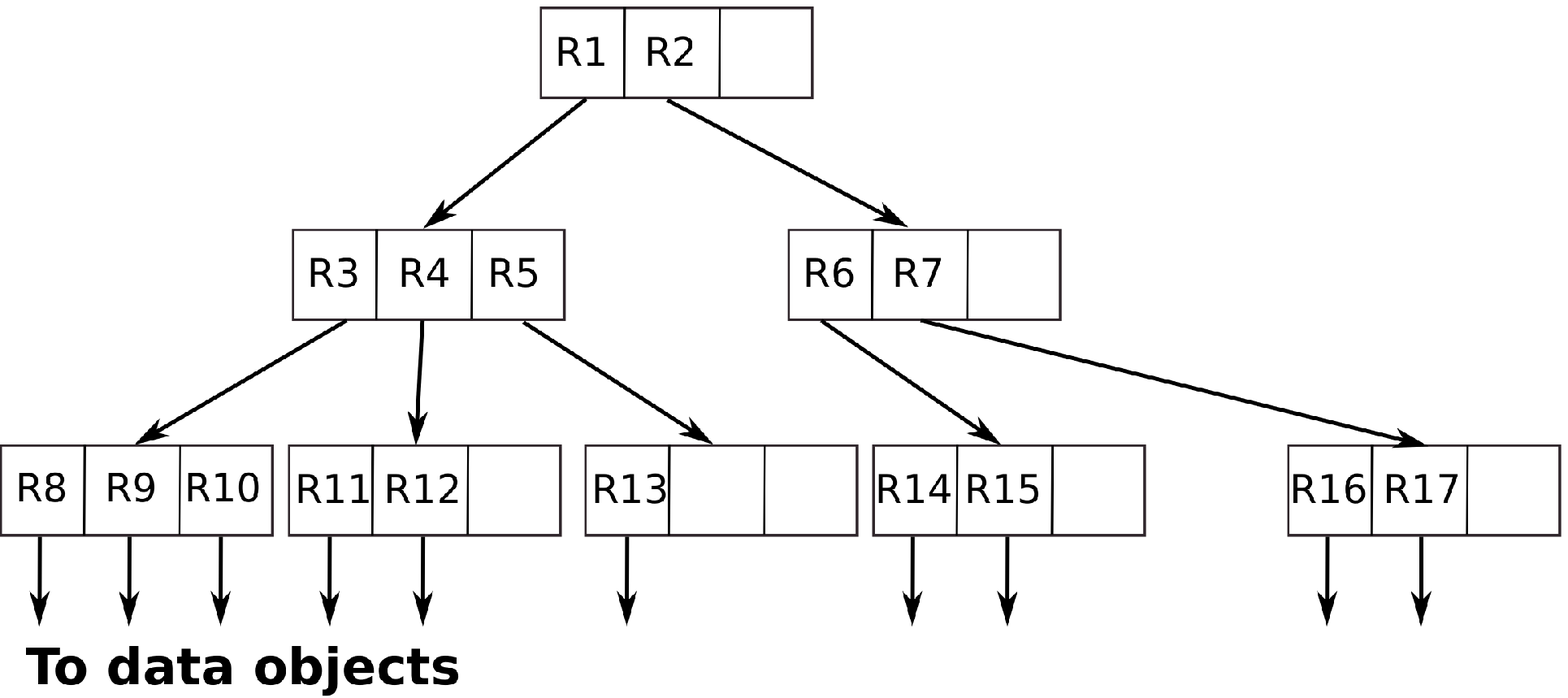}
        \caption[width=.3\textwidth]{}
        \label{fig_rtree}
    \end{subfigure}
    \begin{subfigure}[t]{0.7\columnwidth}
        \centering
        \includegraphics[width=0.9\columnwidth]{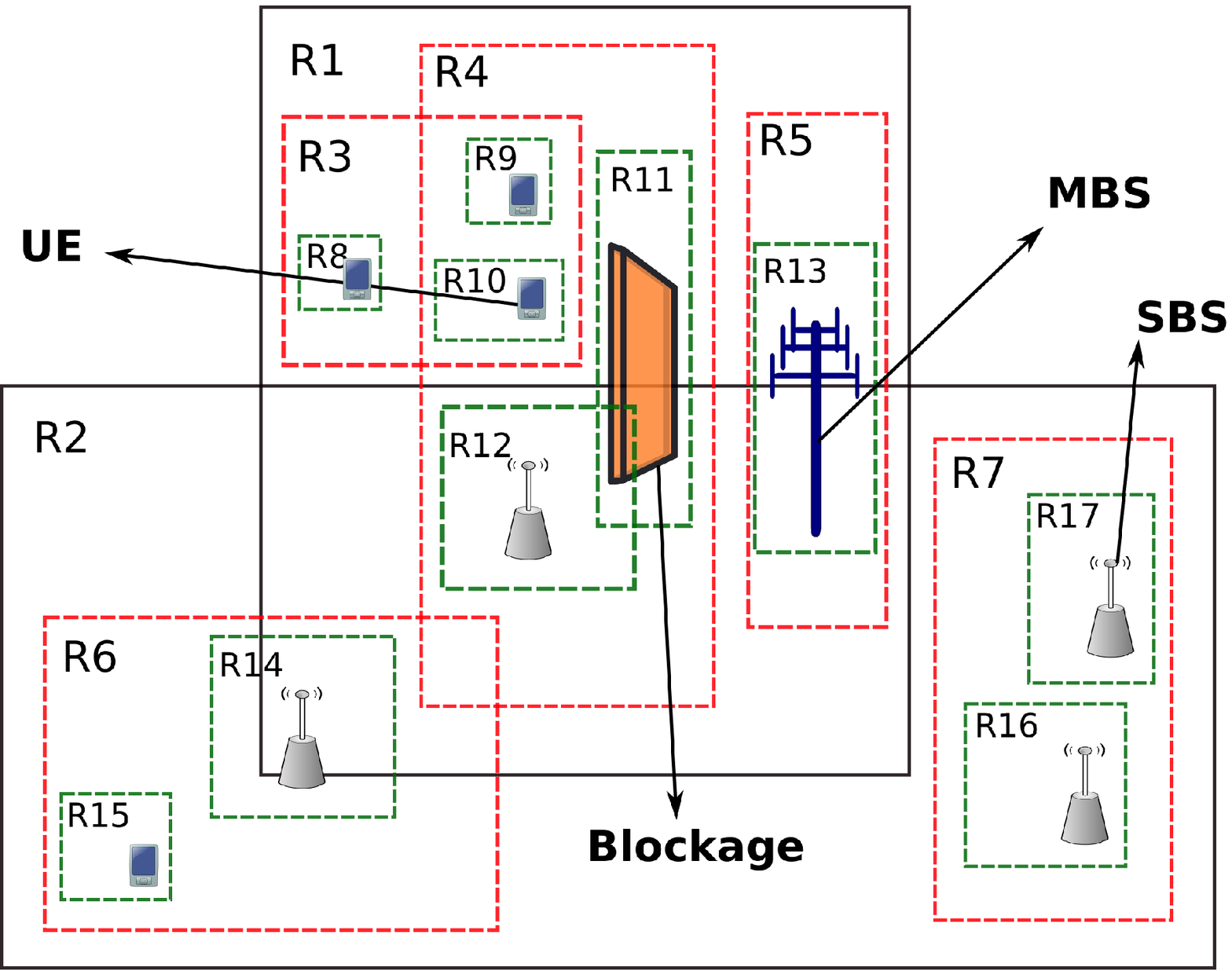}
        \caption[width=.3\textwidth]{}
        \label{fig_simple_net}
    \end{subfigure}
    \caption{\small Example of a HetNet indexed with a single R-tree with $n=10$ and $M=3$. (a) The tree containing different levels of MBRs which partition a network of one macro BS (MBS), four SBSs, four UEs, and one blockage. (b) The rectangles R$1$ to R$17$ represent MBRs, the black MBRs (R$1$, R$2$) are the roots, the red MBRs (R$3$-R$7$) are the second layer, and the green MBRs (R$8$-R$17$) are the leaf nodes which contain the data objects of the network. The MBRs R$1$ and R$2$ cover total area of the network.}
\end{figure*}

\subsection{Architecture}

In order to store heterogeneous nodes on a single geometry tree, nodes of the network need to be represented just by their geometry. In fact, the elements of the network are abstracted as polygons regardless of their higher level nature which can be a UE, BS, or even a blockage. The proposed architecture is shown in Fig.~\ref{fig_hierarchy}. All the elements of the network are generated based on inheritance from an object named \textit{Node}. The \textit{Node} object contains location and geometry (length, width, and height) of the elements and is the lowest level in the platform and is stored on the geometry tree. \textit{Node} is inherited by the \textit{TRX} and the \textit{Blockage} objects. The \textit{TRX} object contains related parameters to a transceiver such as the carrier frequency, transmit power, and antenna properties. Also, the \textit{TRX} contains an independent standard thread with signal-slot capabilities. The signal-slot methods are used to implement message-passing and event-based processes such as asynchronous procedures of the network. Wireless sensors, mmWave or Sub6GHz BSs, and UEs can be generated by inheriting from the \textit{TRX}. The blockage objects are generated by directly inheriting from the \textit{Node}. 
The proposed design consists of a class named \textit{Container} which is used to manage all the created nodes in the simulation. The \textit{Container} holds a geometry tree which indexes all the generated elements and provides the spatial queries over the geometry tree. Since just the Node data is saved on the geometry tree, one single tree can be used for any type of element in the network. The designed architecture and indexing procedure is applicable to any object-oriented language that supports multilevel inheritance. However, the code snippets that we use are based on C++ language.


\subsection{Indexing a HetNet With single Geometry Tree}\label{sec_rtree}


In order to use spatial indexing, a proper spatial data structure should be selected. Most of the spatial data structures work based on the principle of space partitioning and storing data on a tree-like structure such as R-tree or K-d tree~\cite{art_spatial_access}. K-d tree can only contain points and does not handle adding and removing points. However, in R-tree nodes are represented as polygons. Since, we need to provide dimensions for the nodes of the network as well as dynamic removal of them, we use R-tree~\cite{art_Guttman} for spatial indexing. R-tree is a geometry tree proposed by Guttman as a dynamic indexing structure designed for spatial searching in multi-dimensional datasets. Basically, R-tree groups the objects using minimum bounding rectangles (MBR). Objects are added to an MBR within the index that will lead to the smallest increase in its size. R-tree records the indices in its leaf nodes with pointers to the data objects as in Fig.~\ref{fig_rtree}. Data objects refer to the polygons of the elements of the network which is detailed below. Further, by defining $M$ as the maximum number of entries in an element of the tree and $n$ as the size of the HetNet, the average search complexity of R-tree is $\mathcal{O}(\log_M{n})$. A representation of the R-tree and MBRs over a HetNet is illustrated in Fig.~\ref{fig_rtree} and Fig.~\ref{fig_simple_net}, respectively.

\begin{figure}
    \centering
    \begin{subfigure}[t]{0.5\columnwidth}
        \centering
        \includegraphics[width=1\columnwidth]{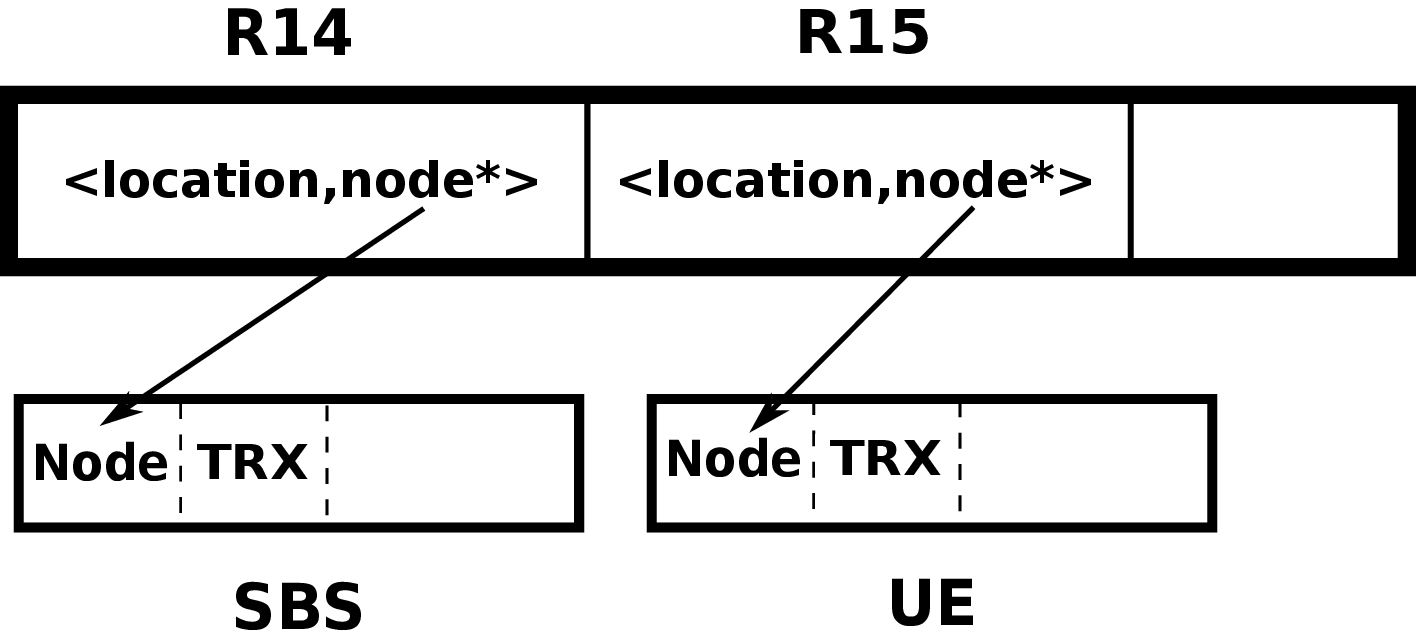}
        \caption[width=.3\textwidth]{}\label{fig_tree}
    \end{subfigure}%
    \begin{subfigure}[t]{0.5\columnwidth}
        \centering
        \includegraphics[width=0.9\columnwidth]{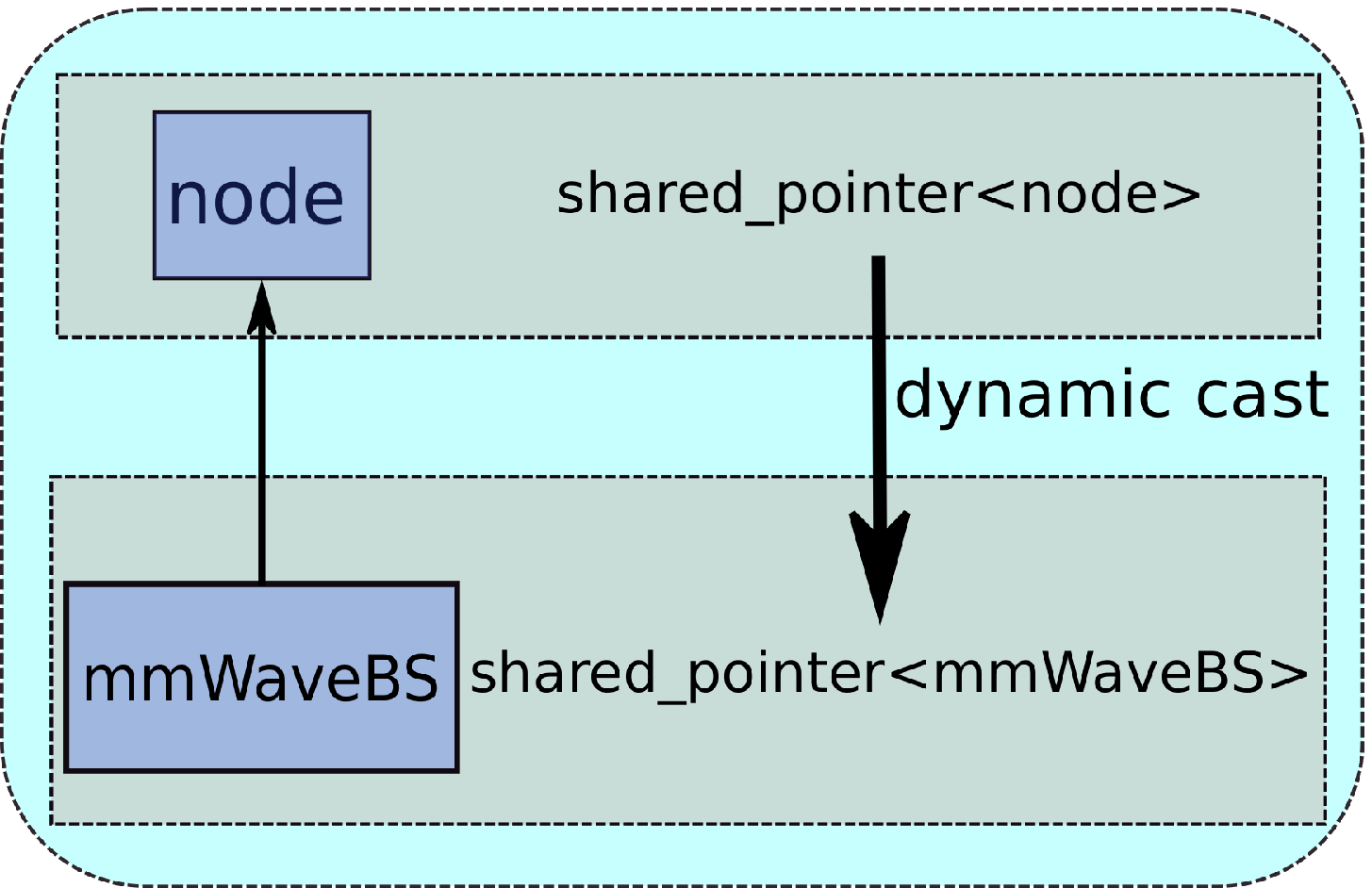}
        \caption[width=.3\textwidth]{}\label{fig_cast}
    \end{subfigure}
    \caption{(a) The \textit{value} pairs of the R-tree leaves and their relationship with the elements of the network. Each leaf contains location and a pointer which stores the \textit{Node} information of the respective element of the network. Here, R$14$ contains the location of a SBS and its \textit{Node} data. (b) Retrieving higher levels of an object from a query result.}
\end{figure}

According to Fig.~\ref{fig_rtree}, the leaf nodes store data objects related to the elements of the network. The data objects are 2-tuples containing first the location of the element, and second a pointer to the \textit{Node} object of the element. We name the 2-tuple, \textit{value} pairs in the code snippets. We define the following \textit{value} pair as the input of the R-tree.
\begin{lstlisting}[style=CStyle]
typedef pair<point,shared_ptr<node>>value;
\end{lstlisting}
According to the above, an element of the network is added to the R-tree based on its location (point variable) and a pointer (shared\_ptr) to its \textit{Node} object and the R-tree indexes the elements of the network based on their corresponding locations. Fig.~\ref{fig_tree} shows the \textit{value} pairs of MBRs R$14$ and R$15$ defined in Fig.~\ref{fig_rtree}. For details of inserting elements to the R-tree see Appendix~\ref{appendix_insert}. 


\subsection{Spatial Queries}\label{sec_query}

A spatial query is a query which considers the spatial relationship between the elements of the network such as relative location-based searching, k-nearest neighbors, and ray tracing. Spatial queries have significant applications in map servers where there are continuous queries over the database based on the location of the objects. Google Maps and Quantum Geographic Information Systems (QGIS) are examples of applications which use spatial queries frequently. Considering the above, any location-dependent search can be defined as a spatial query over the polygons of the elements of the network. For instance, finding fixed-radius neighbors can be defined as a circle-shaped polygon query over the nodes of the network. Therefore, we can represent the association of users to base stations as a spatial query. The same applies to finding the potential interferers in a certain direction which can be stated as a triangular-shaped polygon query. In the following, we describe these queries.

After inserting all the elements (BSs, UEs, blockages) in the R-tree, any element is able to define customized spatial queries over the network. The general format of a spatial query is defined as follows.
\begin{lstlisting}[style=CStyle]
m_tree.query(Condition, Results)
\end{lstlisting}
In the above, the \textit{Condition} can be any function defined based on the \textit{point} variables. \textit{Results} is a standard vector containing the \textit{value} pairs of the elements that their locations satisfy the defined \textit{Condition}. Due to the indexing method, any query over the network results in a vector containing pointers to \textit{Node} objects. In order to derive the higher levels of a \textit{Node} object, for example a mmWave BS from the pointer of its \textit{Node}, we use \textit{dynamic\_cast} as in Fig.~\ref{fig_cast}. It is important to note that since we use downcasting to derive classes from the shared pointer of the \textit{Node}, the \textit{Node} class should be polymorphic~\cite{book_Stroustrup}, i.e., \textit{Node} should at least contain one virtual method.

Here, we use spatial queries to define two common location-dependent searches in wireless networks: search for fixed-radius near neighbors and search for interferer BSs residing in the boresight of a receiver. However, any customized query can be defined as well. The two queries are implemented as follows.

($\rmnum{1}$) Fixed-radius near neighbors: This query is used in association, coordination, and routing problems. The \textit{Condition} for this query is written based on the euclidean distance from the desired point. In fact, any point that is in distance R of the desired point is in a circular polygon with radius R around the desired point. If the MBR of any element of the network intersects with the defined circular polygon, then the element is in distance R of the desired point (center of the circular polygon). In the following the elements located in the defined distance R of the \textit{desired} point are derived.
\begin{lstlisting}[style=CStyle]
// Defining the result vector
std::vector<value>results;
// The desired location.
point desired(xx,yy);
m_tree.query(bgi::satisfies([&](value const& v){return bg::distance(v.first, sought)<R;}),std::back_inserter(results));
\end{lstlisting}

\begin{figure}
    \centering
    \begin{subfigure}[t]{0.45\columnwidth}
        \centering
        \includegraphics[width=0.8\columnwidth]{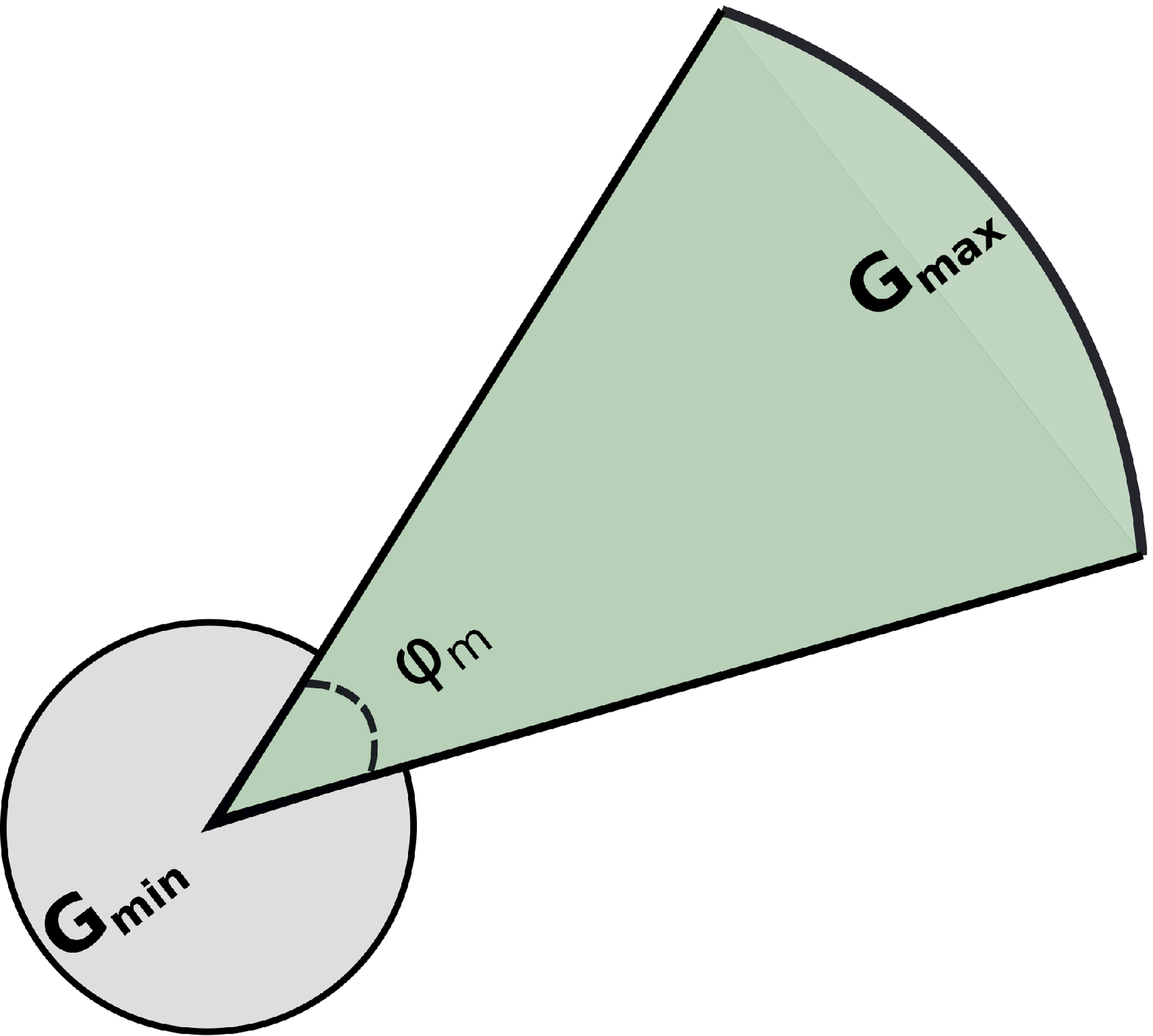}
        \caption{}\label{fig_sectored_pattern}
    \end{subfigure}%
    \begin{subfigure}[t]{0.45\columnwidth}
        \centering
        \includegraphics[width=1\columnwidth]{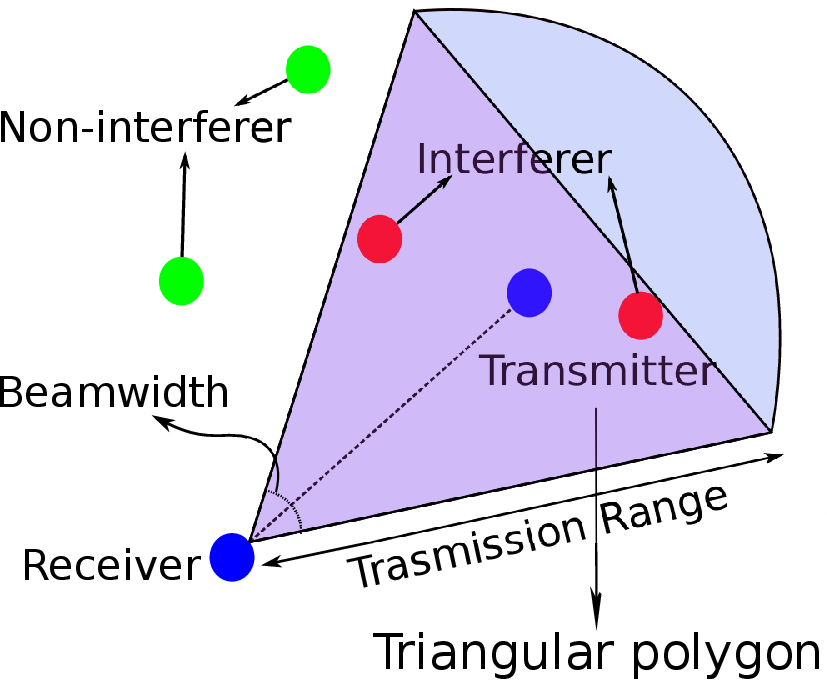}
        \caption{}\label{fig_polygon}
    \end{subfigure}
    \caption{(a) Sectored-pattern antenna model with the beamwidth of $\phi_m$, main-lobe gain of $G_{\text{max}}$, and side-lobe gain of $G_{\text{min}}$. (b) Polygon estimating the area in which potential interferer BSs reside.}
    \end{figure}
    
($\rmnum{2}$) Directional interferer neighbors: This query is used for SINR calculation of a directional wireless link. In another terms, search for neighbors in distance and azimuth (or elevation) at the same time. In directional communications, the power received at the receiver depends on the combined antenna gain of the receiver and transmitter. Directional communication is viable with large antenna arrays and using different MIMO transmission techniques such as fully digital, analog or hybrid beamforming. Here, we use the sectored-pattern antenna gain shown in Fig.~\ref{fig_sectored_pattern} which is practically accepted for single stream analog beamforming~\cite{art_Jeff_0}. In order to accurately calculate the antenna gain at a receiver, we need to figure out if the interfering transmitter is in the main lobe of the receiver or not. We solve this problem with a polygon query. In order to define this query, we define a triangular polygon estimating the area in which the BSs causing interference are located as in Fig.~\ref{fig_polygon}. The nodes residing inside the triangular polygon are in the main lobe of the receiver. After finding the interferer nodes, we can initiate the same query from the interfering nodes to see if the receiver is in their main lobe as well. This query can be implemented as follows.
    
\begin{lstlisting}[style=CStyle]
// Defining the result vector.
std::vector<value> results;
// Performing the query to search for any node intersecting with the derived polygon.
m_tree.query(bgi::intersects(triangular_polygon), std::back_inserter(result));
\end{lstlisting}
In the above, the \textit{triangular\_polygon} is defined for a transmitter-receiver pair based on the direction of transmission, beamwidth, and the maximum transmission range according to Fig.~\ref{fig_polygon}. The neighbors whose MBR intersect with the \textit{triangular\_polygon} are returned in the \textit{results} vector. By using \textit{dynamic\_cast} the \textit{TRX} related object of the neighbors can be derived. Finally, the interference can be calculated based on the interference model of the network. It is worth mentioning that the polygon query can be used for other purposes such as user association for BSs which have multiple sectors as well.

\section{Simulation Acceleration with Spatial Indexing}
\begin{figure*}
    \centering
    \begin{subfigure}[t]{.65\columnwidth}
        \centering
        \includegraphics[width=1\columnwidth]{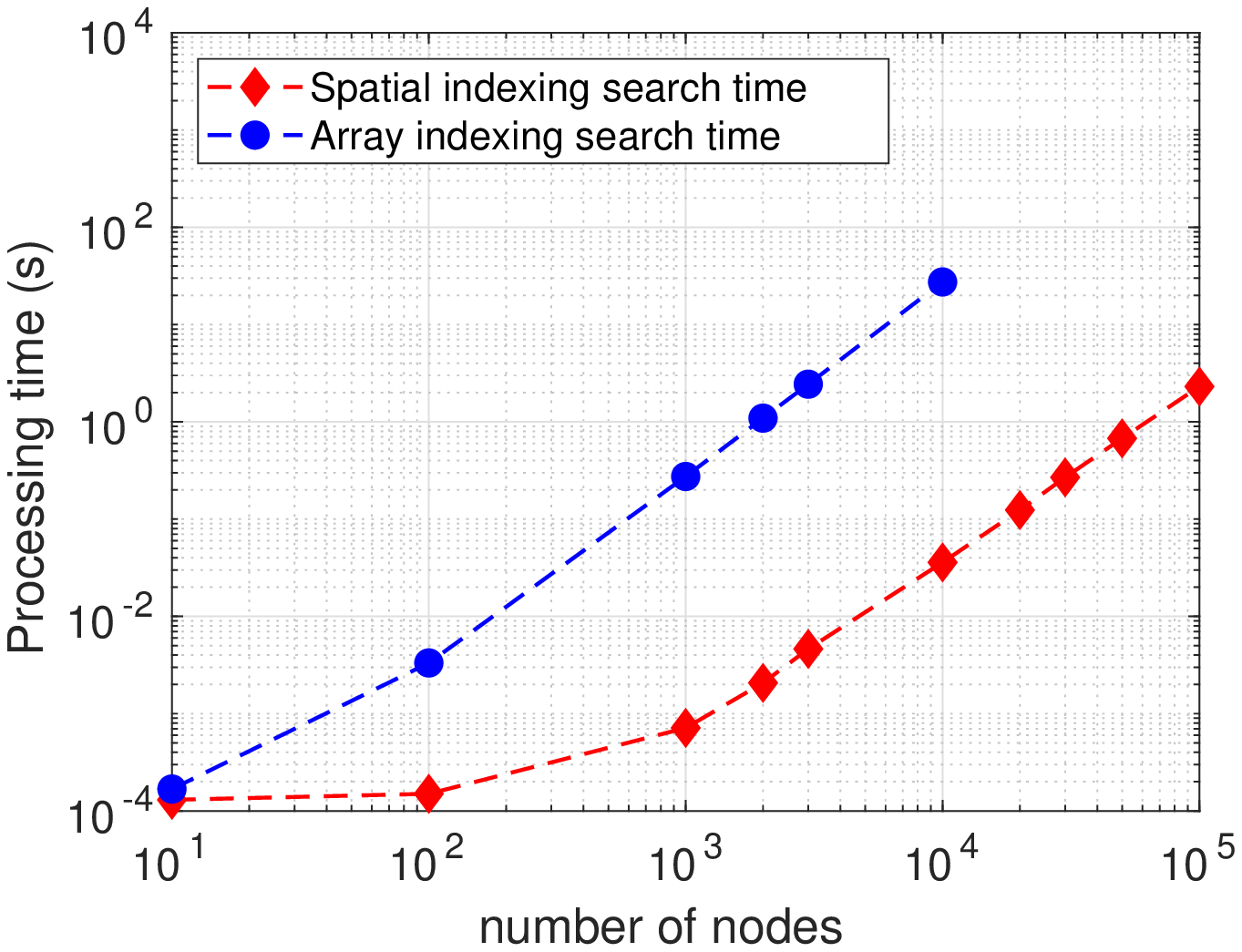}
        \caption{}\label{fig_snr}
    \end{subfigure}%
    \begin{subfigure}[t]{.65\columnwidth}
        \centering
        \includegraphics[width=1\columnwidth]{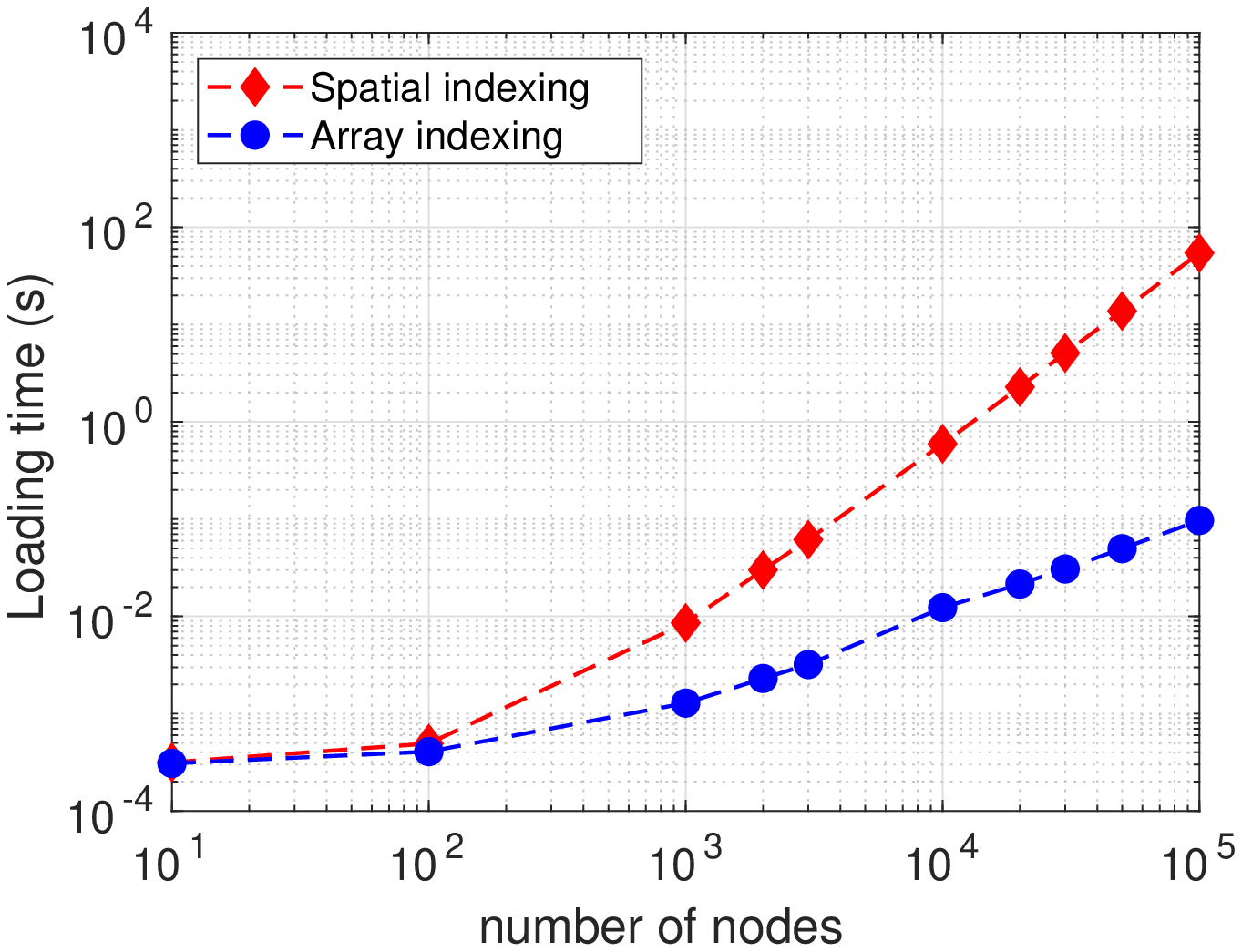}
        \caption{}\label{fig_load}
    \end{subfigure}
    \begin{subfigure}[t]{.65\columnwidth}
        \centering
        \includegraphics[width=1\columnwidth]{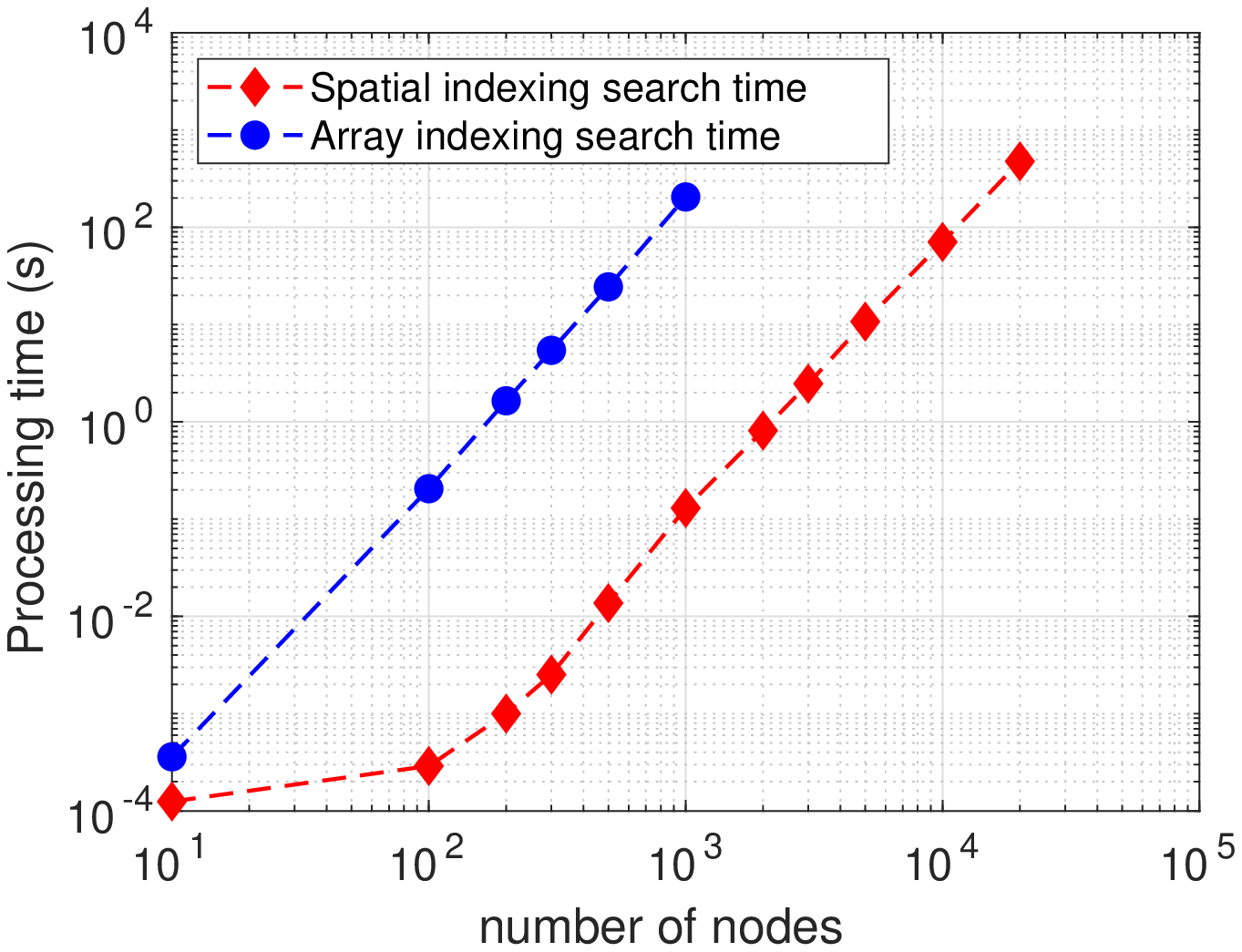}
        \caption{}\label{fig_sinr}
    \end{subfigure}
    \caption{(a) Processing times to calculate SNR of all existing links. (b) Loading time to generate and store all nodes. (c) Processing time to calculate SINR of all existing links. }
\end{figure*}
\vspace{-2pt}

The goal of this section is to compare the performance of spatial indexing versus array indexing in location-dependent searches in large directional wireless networks. The network under study contains mmWave SBSs which are distributed with fixed density of $100~\text{SBS}/\text{Km}^2$. We assume SBSs are equipped with directional antennas. Without loss of generality, we assume all SBSs are on the same horizon plane. Thus, we do not consider beams variations in the elevation and define the antenna gain pattern with widely-adopted sectorized-pattern as~\cite{art_Jeff_0,art_omid_cache}.

Generally, in cellular communication, the measured signal at a receiver is a combination of the desired signal, the interference, and noise, hence SINR. In mmWave communications, due to narrow beams, the links are sometimes assumed to be interference-free and the SNR metric is used in simulations. Despite the fact that the interference-free assumption is reasonable, the probability of occurrence of interference increases as the density of the network increases~\cite{art_niknam, art_SIR_mmWave}. 
Selection of the right metric is important in certain applications such as path selection and routing algorithms. Hence, we assume two scenarios: ($\rmnum{1}$) SNR and ($\rmnum{2}$) SINR calculation of all potential links between any two pair of SBSs with maximum transmission range of $200$~m. 

In the following we provide performance time comparison and complexity analysis in both scenarios. Experiments are carried out in C++ with an Intel(R) Core(TM) i5-6300HQ~@~2.30~GHz processor powered by Fedora $31$ operating system and Linux kernel $5.4.18$.

\vspace{-2pt}
\subsection{SNR calculation}

In order to calculate the SNR of all potential links between SBSs in a network, we need to measure distance of all pairs of SBSs and calculate SNR of the ones in transmission range of each other. In array indexing, complexity of finding potential links between one SBS and its neighbors is $\mathcal{O}(n)$, which contains measuring the distance between the node and all existing nodes. Hence, the total complexity for all the nodes is $\mathcal{O}(n^2)$. However, with spatial indexing, finding neighbors can simply be implemented with a fixed-radius neighbor query as in~\ref{sec_query} which is a spatial query over the distance. The complexity of one query is $\mathcal{O}(\log n)$ and hence for the whole network is $\mathcal{O}(n\log n)$ on average. In Fig.~\ref{fig_snr} the processing time for finding the potential links for calculating SNR of the network is presented. As shown, spatial indexing outperforms array indexing. Further, the processing time for array indexing when $n=10^5$ is cutoff since it is extremely high and pointless to calculate.

It is worth mentioning that, when using spatial indexing, loading the nodes on the R-tree introduces an overhead to the simulation. However, loading time is a \textit{one-time} overhead, however, location-dependent searches are called frequently during a simulation. In Fig.~\ref{fig_load}, we have compared the required time of storing nodes on an array and a R-tree with respect to the number of nodes.


\subsection{SINR calculation}

In directional communication, the calculation of SINR for a link contains one additional search compared to SNR calculation. In fact, after finding the potential neighbors, we need to search for interferers for each link. This search is in distance and azimuth according to Fig.~\ref{fig_polygon}. With array indexing, finding directional interferers for each link leads to another search which increases the complexity to $\mathcal{O}(n^2)$. Hence, the computational complexity for the whole network can go up to $\mathcal{O}(n^3)$. On the other hand, spatial indexing provides a systematic approach to accelerate the calculation of SINR. SINR calculation can be simply implemented as a combination of fixed-radius near neighbors query followed by a triangular polygon query over the results of the first query. This systematic approach is one of the advantages of spatial indexing. In Fig.~\ref{fig_sinr}, the processing time of SINR calculation in large wireless networks with directional communication is plotted. As it is shown in Fig.~\ref{fig_sinr}, spatial indexing has clear advantage in processing time of searches in distance and azimuth. This advantage can be used to enormously accelerate system-simulation of large systems.





\section{Conclusion and Future Work}
In this paper, we propose the use of spatial indexing in system-level evaluation of 5G heterogeneous cellular networks. We introduced an inheritance based polymorphic architecture which enables us to index a wireless heterogeneous network with a single R-tree. This structure enables us to take advantage of spatial queries to accelerate simulation of large-scale directional heterogeneous wireless networks. Researchers can use spatial indexing in their platforms to accelerate system-level simulations enormously. Acceleration can be achieved in any search defined in distance, azimuth or even elevation. Further, due to the ability of considering the blockage, spatial indexing can accelerate system-level simulations which account for the spatial correlation of blocking such as~\cite{art_saha_blockage}. Another main application of spatial indexing could be generating training data sets of accurate SINR values in millimeter-wave communications for machine learning purposes. Further, spatial indexing can accelerate simulation in multiple applications such SINR evaluations in ad-hoc networks, routing, clustering, implementation of self-organizing networks (SONs)~\cite{art_amiri_twc}, and generating communication graphs. We are currently developing an open-source platform based on the introduced structure in Fig.~\ref{fig_hierarchy} which implements some of the applications of spatial indexing in~\cite{GeoNS}.


\appendices
\section{Insertion to the R-tree}\label{appendix_insert}
In order to implement the R-tree, we have used the Boost C++ libraries. 
Further, we define two variables representing the location and the dimensions of an element of the network respectively as \textit{point} and \textit{box} variables. Without loss of generality, we assume \textit{point} has two-dimensions. The \textit{point} variable can be defined for three-dimensional data if the height of the elements of the network is important as well. Also, The \textit{point} variable is defined over \textit{float} data type to accelerate the simulations as follow.
\begin{lstlisting}[style=CStyle]
namespace bg = boost::geometry;
namespace bgm = boost::geometry::model;
typedef bgm::point<float,2,bg::cs::cartesian> point;
typedef bgm::box<point> box;
\end{lstlisting}
The \textit{Node} contains the above definitions as the geometry information of elements of the network as their polygon represented as the \textit{box} variable. As it is mentioned in Section~\ref{sec_rtree}, leaves of the R-tree hold the information as \textit{value} pairs. 
After creating an object, the corresponding \textit{value} pair is created and inserted in the tree. Here, we create a mmWave BS as an example.
\begin{lstlisting}[style=CStyle]
// Generate R-tree 
namespace bgi = boost::geometry::index;
bgi::rtree<value,bgi::quadratic<16>> m_tree;
boost::shared_ptr<mmWaveBS>BS;
// Generate shared pointer of a mmWaveBS
BS=boost::shared_ptr<mmWaveBS>(new mmWaveBS(x,y));
//Insert the BS to the tree.
m_tree.insert(std::make_pair(BS->get_loc(),BS));
\end{lstlisting}
In the first line above, the \textit{m\_tree} is created as an R-tree over the defined \textit{value} pairs. The second and third line create a mmWaveBS and set its location, i.e., $x, y$. Finally, the created mmWaveBS is inserted to the tree with its corresponding \textit{value} in line four (The mmWaveBS contains the \text{get\_loc()} method which returns the location of the object in \textit{point} format.). Also, since mmWaveBS is inherited from the \textit{Node}, there is no need to cast it to the \textit{Node} object.

\bibliographystyle{IEEEtran}
\bibliography{IEEEabrv,Spatial_indexing}
\end{document}